# Compression of CEP-stable multi-mJ laser pulses down to 4 fs in long hollow fibers

**Frederik Böhle[1], Martin Kretschmar[2], Aurélie Jullien[1], Mate Kovacs[3], Miguel Miranda[4], Rosa Romero[5,6], Helder Crespo[6], Uwe Morgner[2,7], Peter Simon[8], Rodrigo Lopez-Martens[1] and Tamas Nagy[2,8]**

[1]Laboratoire d'Optique Appliquée, École Nationale Superieure de Techniques Avancées - Paristech, École Polytechnique, CNRS, 91761 Palaiseau Cedex, France
[2]Institut für Quantenoptik, Leibniz Universität Hannover, Welfengarten 1, 30167 Hannover, Germany
[3]Department of Optics and Quantum Electronics, University of Szeged, P.O. Box 46, Szeged 6701, Hungary
[4]Department of Physics, Lund University, P.O. Box 118, SE-22100 Lund, Sweden
[5]Sphere Ultrafast Photonics, Lda, R. Campo Alegre 1021, Edifício FC6, 4169-007 Porto, Portugal
[6]IFIMUP-IN and Departamento de Física e Astronomia, Universidade do Porto, R. Campo Alegre 687, 4169-007 Porto, Portugal
[7]Laser Zentrum Hannover e.V., Hollerithallee 8, 30419 Hannover, Germany
[8]Laser-Laboratorium Göttingen e.V., Hans-Adolf-Krebs-Weg 1, 37077 Göttingen, Germany

E-mail: nagy@iqo.uni-hannover.de

**Abstract.** CEP-stable 4 fs near-IR pulses with 3 mJ energy were generated by spectral broadening of circularly polarized 8 mJ pulses in a differentially pumped 2-m-long composite stretched flexible hollow fiber. The pulses were characterized using both SHG-FROG and SHG d-scan methods.

Keywords: pulse compression, hollow fiber, few-cycle pulses

PACS: 42.65.-k Nonlinear optics, 42.65.Re Ultrafast processes; optical pulse generation and pulse compression

# 1. Introduction

Few-cycle light pulses have become the most important tools in contemporary ultrafast sciences. Sub-2-cycle carrier-envelope phase (CEP) stabilized pulses are particularly attractive because they provide high temporal resolution in time-resolved experiments and also facilitate the generation of isolated attosecond XUV pulses in a simple way. It is well known that the generated photon number in the XUV spectral range scales with the energy of the driver pulse [1]. Therefore, increasing the driver pulse energy can pave the way for attosecond pump – attosecond probe experiments requiring high XUV yields.

The generation of high-energy sub-2-cycle pulses is a complicated task. Gain narrowing in optically pumped laser media limits the minimal duration of mJ level pulses to ~20 fs. There are two ways to overcome this problem: spectral broadening of the amplified pulses by nonlinear effects or direct amplification of the low-energy pulses by parametric processes. Multi-color pumping of parametric processes allows the amplification of sub-5 fs pulses up to 100 mJ energies as demonstrated in [2]. However, the immense technical challenges connected with this technology currently limit the operation to 10 Hz, which is inconvenient for many applications.

The most commonly used method for spectral broadening of amplified pulses relies on self-phase modulation (SPM) in noble gas-filled hollow-core fibers [3]. In this way, more than octave-spanning spectra at mJ energy level can be obtained allowing pulse compression to sub-5 fs duration [4–10]. In these studies either the output energy did not significantly exceed 1 mJ or no CEP stabilization was achieved [6], which currently limits applications in attosecond science. Very recently, a study has been published concerning the influence of hollow fibers on CEP stability [11], where pulse energies of up to 2.1 mJ at 3.9 fs transform limit were reported at the output of a differentially pumped 1 m long capillary of 320 μm ID. However, neither the output spectrum, nor the achieved energy and duration of the compressed pulses were specified in this study.

In this paper we overview the physical mechanisms relevant for up-scaling the peak power in hollow fiber pulse compressors. By a simple analysis, we show that the effective length as well as the cross-section of the waveguide has to be increased proportionally to the peak power of the input pulse. This requires long capillaries, which are available thanks to an innovative design [12]. For the first time, using a combination of circularly polarized light in a long hollow fiber with a pressure gradient, we demonstrate the generation of 4 fs CEP stabilized pulses with 3 mJ energy. This is the highest pulse energy achieved to date for CEP stabilized sub-5 fs pulses at kHz repetition rate.

# 2. Power scaling of hollow fiber compressors

In order to achieve controlled spectral broadening in the hollow waveguide with minimal losses, two criteria have to be fulfilled [13]: (i) to avoid self-focusing, the peak power ($P$) of the pulse should not exceed the critical power $P_{cr} = \lambda^2/(2\pi n_2)$, and (ii) the peak intensity ($I$) of the pulse should be below the threshold intensity for photo-ionization ($I_{th}$). In the above expression, $\lambda$ denotes the central wavelength and $n_2$ the nonlinear refractive index of the medium. Under these conditions in an ideal waveguide, single-mode operation will be obtained without energy flow between the eigenmodes. The transmission of the setup will be solely determined by the in-coupling efficiency and by the linear propagation losses of the waveguide while the spectral broadening will be exclusively induced by the Kerr-effect, i.e. SPM and self-steepening.

It was shown in [14] that the extent of spectral broadening is almost linearly proportional to the accumulated nonlinear phase, the so called B-integral:

$$B = \frac{2\pi}{\lambda} \int_0^L n_2(z)I(z)dz = \frac{2\pi}{\lambda} n_2 I L_{eff} \qquad (1)$$

Here we introduce the effective interaction length $L_{eff}$, which comprises eventual variations in nonlinear refractive index and intensity along the propagation.

In order to fulfill (i), the peak power limitation sets an upper limit for the intensity of the eigenmode: $I \leq P_{cr}/A_{eff}$, where $A_{eff}$ is the effective mode area of the waveguide. This leads to an upper limit for the B-integral, which depends only on the waveguide geometry:

$$B \leq \frac{2\pi}{\lambda} n_2 \frac{\lambda^2}{2\pi n_2} \frac{1}{A_{eff}} L_{eff} = \frac{\lambda L_{eff}}{A_{eff}} \qquad (2)$$

Furthermore, in order to fulfill (ii), the intensity is limited according to $P \leq I_{th} \cdot A_{eff}$, therefore the effective mode area and also the inner diameter (ID) of the capillary $A_{eff} \sim ID^2$ have to be increased accordingly. In order to scale up the peak power of the pulses while keeping the spectral broadening at the same level, both the inner diameter and the effective length of the waveguide have to be scaled up in the following way:

$$P \sim A_{eff} \sim ID^2 \sim L_{eff} \qquad (3)$$

This scaling law can also be understood in the following way: by increasing the peak power, in order to fulfill (i) the critical power has to be raised accordingly, which results in the decrease of $n_2$. Due to (ii), the peak intensity has to be kept constant at the ionization threshold. The corresponding deficit in the integrand of (1) must therefore be compensated by increasing the integration interval, namely the fiber length.

As discussed above, spectral broadening of pulses at high peak power requires nonlinear media having low nonlinearity together with high ionization threshold. Therefore the most suitable media are helium and neon. Furthermore, in every material the $n_2$ is smaller and the ionization threshold is higher for circularly polarized light than for linear polarized light of the same intensity. Therefore, using circularly polarized light further suppresses the nonlinearities when using high energy pulses [5,15]. It was also shown [5,16] that fluctuations in spectral broadening are significantly reduced by using circularly polarized light.

The scaling law (3) is a result of a rather conservative approach supporting ideal conditions for spectral broadening in hollow fibers, but at multi-mJ energy levels, it becomes hard to fulfill. In real experiments, because of the lack of long capillaries or simply because of limited laboratory space, both conditions are usually overdriven resulting in a sensitive trade-off between overall transmission, beam/pulse quality and overall length. Applying a pressure gradient along the hollow fiber [17] is a powerful technique for preventing self-focusing and filamentation in front of the fiber and maintaining optimal beam coupling into the waveguide. In this case excessive nonlinear interaction takes place only at the end of the waveguide, where it can no longer induce excessive losses [18]. However, the method has a significant drawback: it reduces the effective interaction length by 33%. Therefore, regarding the scaling rule of (3) and the reduction in length of the pressure gradient scheme it is inevitable to use large-aperture and especially long hollow fibers for spectral broadening of high-energy pulses.

## 3. Experiments

In our experiments, 8 mJ 23 fs pulses of a Ti:sapphire high-contrast double-CPA system [19,20] running at 1 kHz were spectrally broadened in long hollow waveguides. The initial linear laser polarization was converted to circular by a $\lambda/4$-waveplate and coupled into the waveguide by telescopic focusing as shown in Fig. 1. The focus position was actively stabilized by a feedback loop to an RMS uncertainty of ~20 µm. After collimating the output of the fiber by a silver mirror, the polarization was converted back to linear using a broadband $\lambda/4$-waveplate (Femtolasers GmbH) and the resulting pulses compressed using by a chirped mirror compressor.

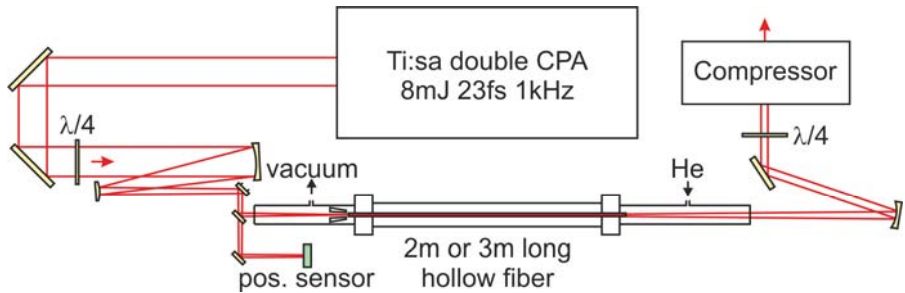

Fig. 1 Experimental arrangement.

The waveguide assembly consists of a few-cm-long solid fused silica taper and a coaxially adjusted stretched flexible hollow fiber of several meters length. The taper filters out the pedestal of the focal distribution, thereby preventing damage at the entrance of the hollow fiber itself. This construction combines the high damage threshold of conventional rigid fibers with the free length scalability and inherent support of the pressure gradient operation of stretched flexible capillaries.

Two different waveguide geometries were considered in the experiments: a 3 meter long fiber with 536 μm ID and a 2 meter long one with 450 μm ID, both having a theoretical transmission of 94% at 790 nm. The distance between the 0.5 mm thick AR-coated fused silica window and the fiber entrance was 97 cm and 136 cm for the longer and smaller fibers, respectively, limited by the available laboratory space of 5.5 m. The respective transmissions of the evacuated waveguides (see Fig. 2) show a significant dependence on the input chirp.

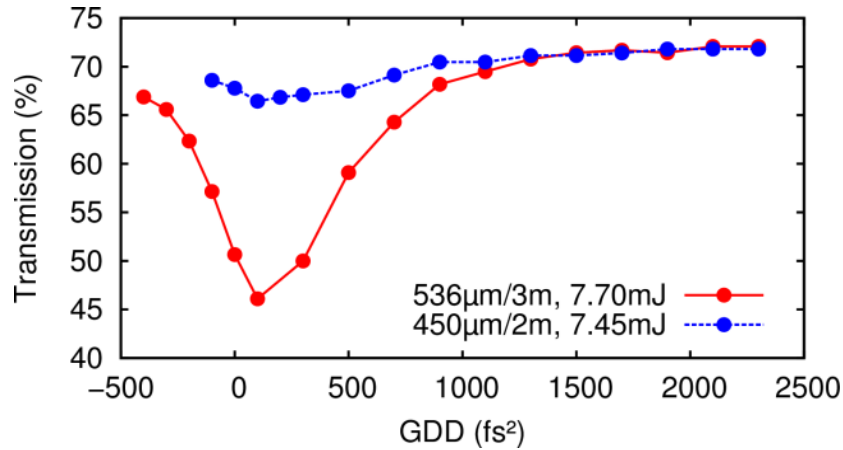

Fig. 2 Transmission of the evacuated fibers as a function of the input chirp.

For large input chirps, the transmission in both cases was as high as 72%, limited by the spatial quality of the input beam. However, near optimal compression, the transmission drops in both cases: it is dramatic in case of the larger fiber and tolerable for the smaller one. This result strongly suggests that the intensity in the entrance window and in air causes significant nonlinear phase distortions. In case of the smaller fiber, the beam size on the window was ~1.65 times larger than for the bigger fiber corresponding to ~2.7 times lower intensity. The estimated B-integrals were thus ~0.37 and ~1 for the smaller and larger waveguides, respectively. It shows how much care has to be taken (especially with large-diameter fibers) when choosing the position of the entrance window.

Spectral broadening of compressed 23 fs pulses (0 fs$^2$) was carried out in the 2-m-long 450 μm hollow fiber. Neither neon nor static gas pressure could be used with this fiber without a substantial drop in transmission. Experiments were carried out with both linear and circular polarizations. The use of circular polarization increased the transmission by ~20% at similar spectral broadening. Furthermore, the RMS spectral fluctuations were also reduced from 7.8% (linear) to 4.7% (circular polarization) as measured for 100 consecutive shots with an integration time of 10 ms. When the waveguide was filled with 1.8 bar helium in pressure gradient mode, regular over-octave-spanning spectra could be achieved (see Fig. 3) with output pulse energies of 3.4 mJ and a circularly symmetric beam profile.

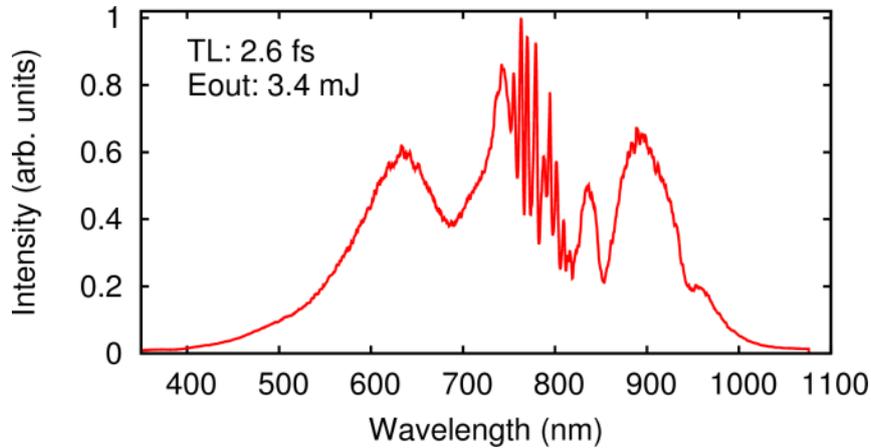

Fig. 3 Output spectrum.

The spectral bandwidth supports a transform-limited pulse duration of 2.6 fs, corresponding to the duration of a single optical cycle at the central wavelength of 742 nm.

During the experiments the CEP drift was measured with a home-made f-to-2f interferometer integrating 4 consecutive pulses. The CEP was stabilized at an RMS error level of 250 mrad before and 360 mrad after the hollow fiber, as shown in Fig. 4.

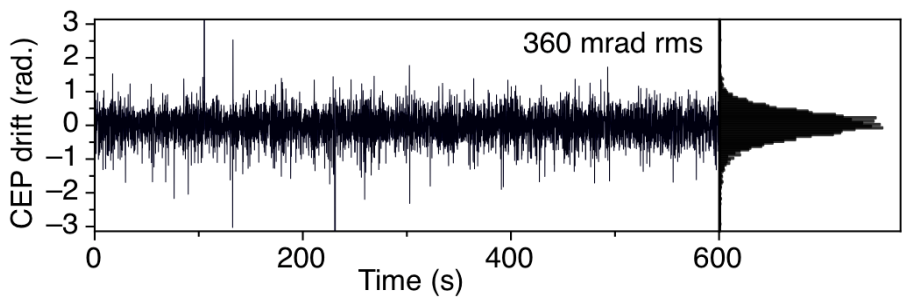

Fig. 4 CEP stability at the output of the fiber.

The CEP noise before the hollow fiber originates from the complex architecture of the employed laser system (double-CPA including a nonlinear temporal filter). Then, the slight deterioration of the CEP stability in the hollow fiber is attributed to the non-zero level of photo-ionization inside the fiber [21] and to fluctuations induced by the finite pointing stability of the laser at the fiber entrance (20 μm RMS).

The pulses were compressed using a set of 12 double-angle chirped mirrors (Ultrafast Innovations) introducing a total of -500 $fs^2$ and a fused silica wedge pair for fine adjustment.

The compressed pulses of 3 mJ energy were first characterized using a home-made single-shot SHG FROG device incorporating a 5 μm thick BBO crystal [22]. The measurement displayed in Fig. 5 yields a pulse duration of 4.27 fs.

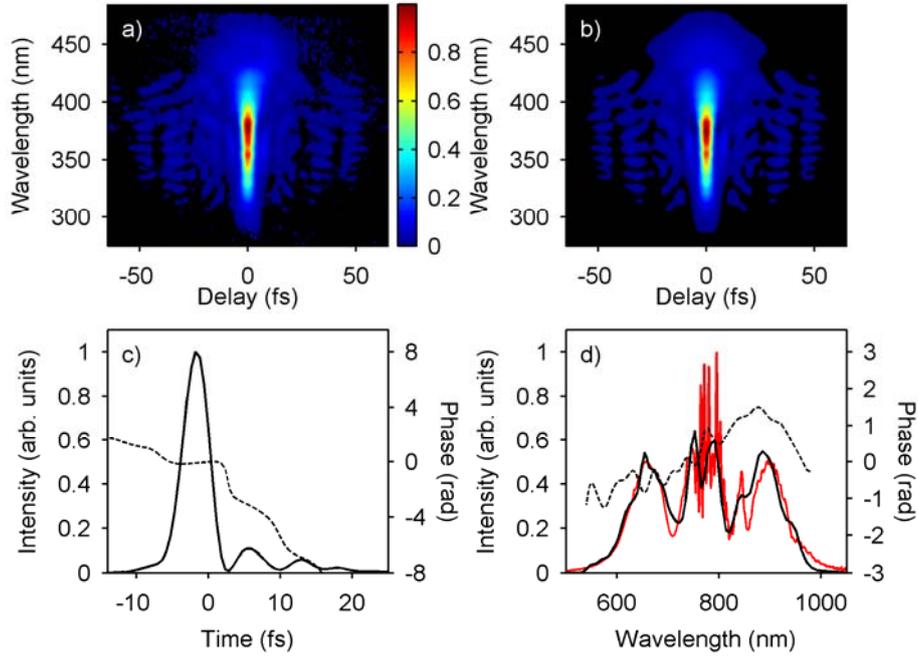

Fig. 5 SHG FROG measurement. Measured (a) and retrieved (b) traces with the retrieved temporal (c) and spectral shapes (d) are shown. The dashed lines represent the phases, the red one depicts the measured spectrum.

The retrieval was carried out on a grid size of 512 resulting in a FROG error of 0.23%. The retrieved spectrum matches well to the experimental one measured directly in the input aperture of the FROG device but it is considerably narrower than the spectrum at the output of the fiber (Fig. 3). The reason is that the single-shot FROG requires a flat top beam profile, therefore only a small portion of the Gaussian-like beam with nearly constant energy distribution was used for the FROG measurement. On the other hand, at the end of the hollow fiber, all spectral components emerge from the same fiber mode with different divergence. The beam profile therefore becomes spectrally inhomogeneous: the blue part of the spectrum will be more concentrated in the core while the red components are distributed over a larger mode area. This effect can clearly be seen on the spatially resolved spectra in [23]. Therefore, sampling a small portion of the beam profile leads to the spectral clipping seen in Fig. 5.

In order to characterize the pulses more accurately, the full beam needs to be focused into the nonlinear medium of the measurement device. The experimental simplicity of the d-scan method [24] combined with its performance in sub-two-cycle pulse measurement [25,26] makes it most suited to this requirement. The setup was similar to the one in [25] albeit with a 5 μm thick BBO crystal [26]. Compression optimization and dispersion scanning was achieved with fused silica wedges. Fig. 6 shows the measured (calibrated) and retrieved d-scan traces. The d-scan retrieval algorithm was run five times with slightly different initial conditions. The shortest pulse duration was $3.9 \pm 0.1$ fs FWHM obtained for the reference glass insertion (denoted as zero in the plots).

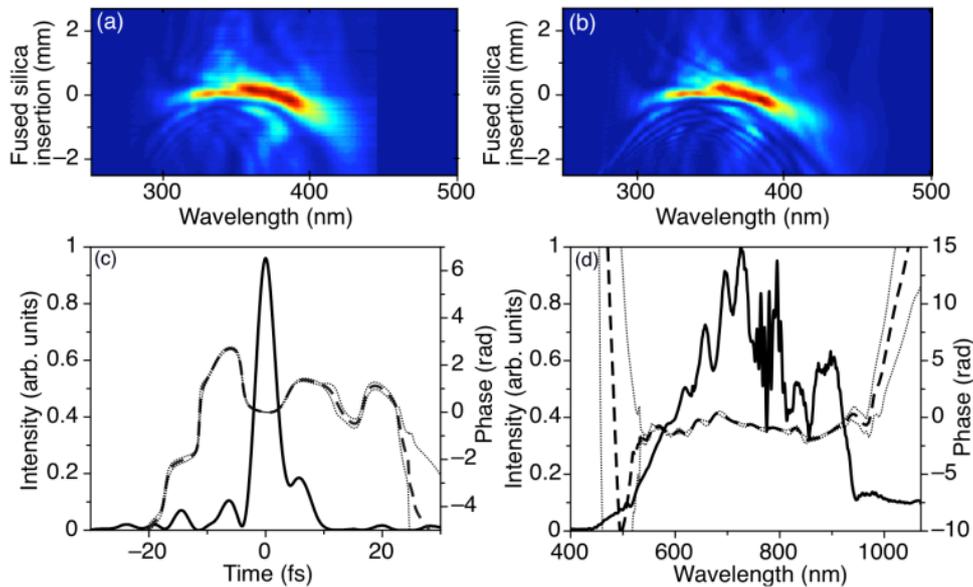

Fig. 6 SHG d-scan measurement. Measured (a) and retrieved (b) traces with the retrieved temporal (c) and spectral shapes (d) are shown. In (c) and (d) dashed lines represent the phase, dotted ones their RMS deviations.

**4. Conclusion**

In conclusion, we showed that increasing the peak power in hollow fiber compressors inevitably requires longer waveguides with larger inner diameters. It is the first demonstration that the stretched flexible hollow fiber technology, offering free length scalability, is applied to the compression of multi-mJ pulses at multi-W average power. By using a 2-meter long 450 µm diameter composite hollow fiber in pressure gradient mode combined with circularly polarized input pulses for the first time, CEP-stable 4 fs pulses corresponding to 1.6 optical cycles were generated with 3 mJ energy resulting in a peak power of 0.7 TW. The transform limited pulse duration of the generated spectrum is 2.6 fs, corresponding to a single optical cycle. This bears the potential for achieving even shorter pulses by further optimizing the dispersion management. The present results represent an energy boost of a factor ~1.9 for CEP stabilized pulses compressed down to sub-5 fs duration at 1 kHz repetition rate and of ~2.5 for 4 fs duration compared to former achievements. Even higher energies are expected by mitigating nonlinearities at the fiber input and by using longer stretched flexible fibers with larger diameters. Furthermore, these pulses feature high temporal contrast thanks to the double-CPA architecture of the seed laser [19] and thus provide an ideal tool for relativistic optics on solid targets [27].

**Acknowledgement**

P. S. and T. N. thank Ferenc Krausz (MPQ Garching) for his advices which helped making the stretched flexible hollow fiber technology suitable for the compression of high-energy pulses at multi-watt average power levels. M. M. acknowledges the support of the Knut and Alice Wallenberg Foundation and the European Research Council (ALMA-227906). H. C. acknowledges grant PTDC/FIS/122511/2010 from Fundação para a Ciência e Tecnologia, Portugal, co-funded by COMPETE and FEDER. Support from the European Science Foundation through the SILMI programme is gratefully acknowledged. This work was supported by LASERLAB-EUROPE (grant agreement no. 284464, EC's Seventh Framework Programme), by the Deutsche Forschungsgemeinschaft (Mo850/16-1) and by Agence Nationale pour la Recherche (ANR-11-EQPX-005-ATTOLAB) and by ASTRE 2010 programme of the Conseil Général de l'Essonne.

**References**
[1] Hergott J-F, Kovacev M, Merdji H, Hubert C, Mairesse Y, Jean E, Breger P, Agostini P, Carré B and Salières P 2002 Extreme-ultraviolet high-order harmonic pulses in the microjoule range Phys. Rev. A **66** 021801
[2] Veisz L, Rivas D, Marcus G, Gu X, Cardenas D, Mikhailova J, Buck A, Wittmann T, Sears C M S, Chou S-W, Xu J, Ma G, Herrmann D, Razskazovskaya O, Pervak V and Krausz F 2013 Generation and Applications of Sub-5-fs Multi-10-TW Light Pulses CLEO Pacific Rim (Kyoto) pp TuD2–3
[3] Nisoli M, DeSilvestri S and Svelto O 1996 Generation of high energy 10 fs pulses by a new pulse compression technique Appl. Phys. Lett. **68** 2793–5
[4] Park J, Lee J and Nam C H 2009 Generation of 1.5 cycle 0.3 TW laser pulses using a hollow-fiber pulse compressor Opt. Lett. **34** 2342
[5] Chen X W, Jullien A, Malvache A, Canova L, Borot A, Trisorio A, Durfee C G and Lopez-Martens R 2009 Generation of 4.3 fs, 1 mJ laser pulses via compression of circularly polarized pulses in a gas-filled hollow-core fiber Opt. Lett. **34** 1588–90
[6] Bohman S, Suda A, Kanai T, Yamaguchi S and Midorikawa K 2010 Generation of 5.0 fs, 5.0 mJ pulses at 1kHz using hollow-fiber pulse compression. Opt. Lett. **35** 1887–9
[7] Wirth A, Hassan M T, Grguras I, Gagnon J, Moulet A, Luu T T, Pabst S, Santra R, Alahmed Z A, Azzeer A M, Yakovlev V S, Pervak V, Krausz F and Goulielmakis E 2011 Synthesized light transients. Science **334** 195–200
[8] Chen X, Malvache A, Ricci A, Jullien A and Lopez-Martens R 2011 Efficient hollow fiber compression scheme for generating multi-mJ, carrier-envelope phase stable, sub-5 fs pulses Laser Phys. **21** 198–201
[9] Frank F, Arrell C, Witting T, Okell W A, McKenna J, Robinson J S, Haworth C A, Austin D, Teng H, Walmsley I A, Marangos J P and Tisch J W G 2012 Invited review article: technology for attosecond science. Rev. Sci. Instrum. **83** 071101
[10] Schweinberger W, Sommer A, Bothschafter E, Li J, Krausz F, Kienberger R and Schultze M 2012 Waveform-controlled near-single-cycle milli-joule laser pulses generate sub-10 nm extreme ultraviolet continua. Opt. Lett. **37** 3573–5
[11] Lücking F, Trabattoni A, Anumula S, Sansone G, Calegari F, Nisoli M, Oksenhendler T and Tempea G 2014 In situ measurement of nonlinear carrier-envelope phase changes in hollow fiber compression Opt. Lett. **39** 2302
[12] Nagy T, Forster M and Simon P 2008 Flexible hollow fiber for pulse compressors Appl. Opt. **47** 3264–8
[13] Vozzi C, Nisoli M, Sansone G, Stagira S and De Silvestri S 2005 Optimal spectral broadening in hollow-fiber compressor systems Appl. Phys. B **80** 285–9
[14] Pinault S C and Potasek M J 1985 Frequency Broadening by Self-Phase Modulation in Optical Fibers J. Opt. Soc. Am. B **2** 1318–9
[15] Ghimire S, Shan B, Wang C and Chang Z 2005 High-energy 6.2-fs pulses for attosecond pulse generation Laser Phys. **15** 838–42
[16] Malvache A, Chen X, Durfee C G, Jullien A and Lopez-Martens R 2011 Multi-mJ pulse compression in hollow fibers using circular polarization Appl. Phys. B **104** 5–9
[17] Suda A, Hatayama M, Nagasaka K and Midorikawa K 2005 Generation of sub-10-fs, 5-mJ-optical pulses using a hollow fiber with a pressure gradient Appl. Phys. Lett. **86** 111116
[18] Nurhuda M, Suda A, Midorikawa K, Hatayama M and Nagasaka K 2003 Propagation dynamics of femtosecond laser pulses in a hollow fiber filled with argon: constant gas pressure versus differential gas pressure J. Opt. Soc. Am. B **20** 2002–11
[19] Jullien A, Ricci A, Böhle F, Rousseau J-P, Grabrielle S, Forget N, Jacqmin H, Mercier B and Lopez-Martens R 2014 Carrier-envelope phase stable, high-contrast, double-CPA laser system *to be published in* Opt. Lett.
[20] Ricci A, Jullien A, Rousseau J P and Lopez-Martens R 2013 Front-end light source for a waveform-controlled high-contrast few-cycle laser system for high-repetition rate relativistic optics Appl. Sci. **3** 314

[21]     Okell W A, Witting T, Fabris D, Austin D, Bocoum M, Frank F, Ricci A, Jullien A, Walke D, Marangos J P, Lopez-Martens R and Tisch J W G 2013 Carrier-envelope phase stability of hollow fibers used for high-energy few-cycle pulse generation. Opt. Lett. **38** 3918–21

[22]     Akturk S, D'Amico C and Mysyrowicz A 2008 Measuring ultrashort pulses in the single-cycle regime using frequency-resolved optical gating J. Opt. Soc. Am. B **25** A63–A69

[23]     Nagy T, Pervak V and Simon P 2011 Optimal pulse compression in long hollow fibers Opt. Lett. **36** 4422–4

[24]     Miranda M, Fordell T, Arnold C, L'Huillier A and Crespo H 2012 Simultaneous compression and characterization of ultrashort laser pulses using chirped mirrors and glass wedges. Opt. Express **20** 688–97

[25]     Miranda M, Arnold C L, Fordell T, Silva F, Alonso B, Weigand R, L'Huillier A and Crespo H 2012 Characterization of broadband few-cycle laser pulses with the d-scan technique. Opt. Express **20** 18732–43

[26]     Silva F, Miranda M, Alonso B, Rauschenberger J, Pervak V and Crespo H 2014 Simultaneous compression, characterization and phase stabilization of GW-level 14 cycle VIS-NIR femtosecond pulses using a single dispersion-scan setup Opt. Express **22** 10181–90

[27]     Thaury C, Quere F, Geindre J P, Levy A, Ceccotti T, Monot P, Bougeard M, Reau F, D'Oliveira P, Audebert P, Marjoribanks R and Martin P H 2007 Plasma mirrors for ultrahigh-intensity optics Nat. Phys. **3** 424–9

**Figure captions**

Fig. 1 Experimental arrangement.

Fig. 2 Transmission of the evacuated fibers as a function of the input chirp.

Fig. 3 Output spectrum.

Fig. 4 CEP stability at the output of the fiber.

Fig. 5 SHG FROG measurement. Measured (a) and retrieved (b) traces with the retrieved temporal (c) and spectral shapes (d) are shown. The dashed lines represent the phases, the red one depicts the measured spectrum.

Fig. 6 SHG d-scan measurement. Measured (a) and retrieved (b) traces with the retrieved temporal (c) and spectral shapes (d) are shown. In (c) and (d) dashed lines represent the phase, dotted ones their RMS deviations.